\documentclass[twocolumn,aps,prb,superscriptaddress]{revtex4-2}
\usepackage{graphicx}
\usepackage{bm}
\usepackage[linktocpage=true,colorlinks=true,urlcolor=blue,linkcolor=red,citecolor = blue]{hyperref}
\usepackage{pgfplots}
\usepackage{amsfonts}
\usepackage{amsmath}
\usepackage{amssymb}
\usepackage{xcolor}
\usepackage{braket}
\usepackage{relsize}
\usetikzlibrary{arrows}
\usepackage{enumerate}
\usepackage{pgfplots}
\usepackage{newfloat}
\usepackage{siunitx}
\usepackage{physics}
\usepackage{mathtools}
\usepackage{comment}
\usepackage{soul}
\usepackage[T1]{fontenc}
\usetikzlibrary {arrows.meta} 
\usepackage{times}
\usepackage{esint}

\newcommand{\red}[1]{\textcolor{black}{#1}}

\newcommand{\BS}[1]{\boldsymbol{#1}}
\newcommand{\T}[1]{\text{#1}}

\DeclareFloatingEnvironment[name={Supplementary Figure}]{suppfigure}

\graphicspath{ {Figures/} }

\usepackage{subfiles}

\begin{document}

\title{Statistical characterization of valley coupling in Si/SiGe quantum dots via $g$-factor measurements near a valley vortex}
\author{Benjamin D. Woods}
\affiliation{Department of Physics, University of Wisconsin-Madison, Madison, WI 53706, USA}
\author{Merritt P. Losert}
\affiliation{Department of Physics, University of Wisconsin-Madison, Madison, WI 53706, USA}
\author{Nasir R. Elston}
\affiliation{Department of Physics, University of Wisconsin-Madison, Madison, WI 53706, USA}
\author{Hudaiba Soomro}
\affiliation{Department of Physics, University of Wisconsin-Madison, Madison, WI 53706, USA}
\author{M. A. Eriksson}
\affiliation{Department of Physics, University of Wisconsin-Madison, Madison, WI 53706, USA}
\author{S. N. Coppersmith}
\affiliation{School of Physics, University of New South Wales, Sydney, New South Wales 2052, Australia}
\author{Robert Joynt}
\affiliation{Department of Physics, University of Wisconsin-Madison, Madison, WI 53706, USA}
\author{Mark Friesen}
\affiliation{Department of Physics, University of Wisconsin-Madison, Madison, WI 53706, USA}

\begin{abstract}
The presence of low-energy valley excitations in Si/SiGe heterostructures often causes spin qubits to fail. 
It is therefore important to develop robust protocols for characterizing the valley coupling. 
Here, we show that realistically sized samplings of valley energy distributions tend to dramatically overestimate the average valley coupling. 
But we find that knowledge of the valley phase, in addition to the valley splitting, can significantly improve our estimates. 
Motivated by this understanding, we propose a novel method to probe the valley phase landscape across the quantum well using simple $g$-factor measurements. 
An important calibration step in this procedure is to measure $g$ in a loop enclosing a valley vortex, where the valley phase winds by $\pm 2\pi$ around a zero of the valley splitting. 
This proposal establishes an important new tool for probing spin qubits, and it can be implemented in current electron shuttling experiments.
\end{abstract}

\maketitle

Spin qubits in Si/SiGe quantum dots have made tremendous progress towards quantum computing at scale~\cite{Burkard2023,Xue2022,Mills2021,Noiri2022,Philips2022}. 
However, the valley degree of freedom arising from the conduction-band valley degeneracy of Si \cite{Zwanenburg2013} presents a serious challenge, as small energy splittings $E_v$ between the ground and excited valley states can cause decoherence~\cite{Buterakos2021,Teske2023,Langrock2023} (e.g., accidental excitations), and can interfere with state preparation and measurement~\cite{Tagliaferri2018,Weinstein2023}.
While there have been encouraging experimental improvements in increasing the average $E_v$ \cite{Chen2021,McJunkin2021,Wuetz2021,Esposti2024,Volmer2024,Stehouwer2025,George2025}, eliminating all low-$E_v$ dots remains an outstanding challenge.
Indeed, due to the unavoidable presence of alloy disorder in the SiGe regions of the heterostructure, $E_v$ can vary widely from dot to dot~\cite{Wuetz2021}.
Moreover, recent theoretical work~\cite{Losert2023,Lima2023b,Lima2023a,Pena2024} suggests that most, if not all, modern Si/SiGe quantum wells should fall into the ``disorder-dominated'' regime, in which a significant fraction of dots are guaranteed to have low $E_v$.
While this issue may be sidestepped in experiments involving just a handful of dots, a large-scale quantum computer will unavoidably suffer from compromised or failed qubits.
An important goal is therefore to engineer Si/SiGe quantum wells with ``deterministically enhanced'' valley splittings, for which the occurrences of low-$E_v$ dots are exponentially suppressed.
To establish such deterministic behavior, we need to determine the statistical parameters $\Delta_0$ and $\sigma_\Delta$ that characterize the valley coupling distribution~\cite{Losert2023}. 
Here, $E_v = 2|\Delta|$, where the complex, inter-valley coupling $\Delta = \Delta_0 + \Delta_{\delta}$ can be decomposed into its deterministic ($\Delta_0$) and random ($\Delta_{\delta})$ components. 
We note that $\Delta_0$ is determined by the average Ge concentration profile (i.e., the virtual-crystal potential), while 
$\Delta_{\delta}$ is determined by alloy disorder and follows a complex normal distribution with standard deviation $\sigma_{\Delta}$.

To help clarify this situation, below we critically examine the standard method, namely maximum likelihood estimation, for characterizing valley splitting distributions.
We focus particularly on the question of whether this method can reliably determine the parameters $\Delta_0$ and $\sigma_\Delta$ by considering $E_v$ measurements alone.
We show that, in the disorder-dominated regime (defined as $|\Delta_0|/\sigma_\Delta\lesssim\sqrt{\pi}/2$), valley splitting distributions can 
sufficiently mimic deterministically enhanced distributions such that differentiating them with realistically sized data sets is fraught with difficulty.
To clarify this situation, we propose to use measurements of the valley phase $\phi_v$~\cite{Friesen2007}, defined as $e^{i \phi_v}=\Delta / |\Delta|$, in addition to $E_v$.
The valley phase is of interest for fundamental reasons~\cite{Tagliaferri2018,Buterakos2021,Tariq2022}, and it also provides a sensitive probe of interface and alloy disorder~\cite{Losert2023}.

Unfortunately, $\phi_v$ cannot be probed directly~\cite{Zhao2022,Borjans2021}, so a robust measurement scheme is needed. 
Here, we propose to map out $\phi_v$ fluctuations by measuring the Land\'e $g$-factor.
Such $g$-factor measurements are convenient and accurate when performed with qubit resonance experiments.
The mapping relies upon a direct but largely unexplored relationship between $\phi_v$ and $g$, which we describe in detail below.
To summarize: the variations of $g$ across a sample are proportional to $\cos\phi_v$, thus taking extreme values whenever $\phi_v=0$ or $\pi$, as governed by random-alloy fluctuations.
To complete the mapping between $g$ and $\phi_v$, we must therefore apply a calibration step to identify the extrema of the $g$-factor.
We propose to do this by first locating a valley vortex (VV), defined as one of the frequent points where $E_v = 0$.
By moving a single-electron dot in a loop around the vortex, the valley phase winds between 0 and $\pm 2 \pi$, as required.
Indeed, a recent shuttling experiment has demonstrated that $E_v$ can be mapped out over large regions of heterostructures~\cite{Volmer2024}; here, we suggest using a similar technique to map out the $g$-factor and to perform the calibration step.


\textbf{Characterizing the valley coupling distributions.~}--~
Knowledge of the valley splitting distribution allows us to predict the failure rate of qubits at scale.
We first consider the $\Delta$ distribution, for which the real and imaginary components, $\Delta_r$ and $\Delta_i$, are uncorrelated and normally distributed, having a probability density of
\begin{multline}
    f(\Delta ; |\Delta_0|,\phi_0, \sigma_\Delta) = (\pi \sigma_{\Delta}^2)^{-1}
    \\ \times
    e^{-
    \left(\Delta_r - |\Delta_0|\cos\phi_0\right)^2
    /\sigma_{\Delta}^2
    }
    e^{-
    \left(\Delta_i - |\Delta_0|\sin\phi_0\right)^2
    /\sigma_{\Delta}^2
    }. \label{fDelta} 
\end{multline}
The distribution is determined by the average Ge concentration profile of the quantum well, the vertical electric field, and the lateral shape of the quantum dot, and is fully characterized by the parameters $|\Delta_0|,\phi_0$, and  $\sigma_\Delta$, where $\phi_0$ is the average valley phase defined in $\Delta_0=|\Delta_0|e^{i\phi_0}$.
The deterministic valley coupling $\Delta_0$ is mainly determined by the shape of the vertical confinement potential~\cite{Losert2023, Friesen2007}, particularly its sharpest features, while its standard deviation $\sigma_\Delta$ is fully determined by the weighted overlap of the quantum dot wave function with Ge atoms~\cite{Losert2023}.
Since $E_v=2|\Delta|$ is much easier to measure that $\Delta$, we also consider the distribution for $E_v$, which has a Rician form~\cite{Losert2023} given by
\begin{equation}
    f(E_v ; |\Delta_0|,\sigma_\Delta) = \frac{E_v}{2 \sigma_{\Delta}^2}
    e^{-(E_v^2 + 4|\Delta_0|^2)/(4 \sigma_\Delta^2)}
    I_0\left(\frac{E_v |\Delta_0|}{\sigma_\Delta^2}\right), \label{EvDis}
\end{equation}
where $I_0(x)$ is a modified Bessel function of the first kind.
Scaling $E_v$ by $\sigma_\Delta$ in this expression, we see that the shape of the distribution is entirely determined by the ratio $|\Delta_0|/\sigma_\Delta$, with a crossover to deterministically enhanced behavior occurring when $|\Delta_0|/\sigma_\Delta > \sqrt{\pi}/2\approx 0.9$.
A natural strategy for characterizing $f(E_v)$ is to gather statistics on $E_v$, either from many dots on the same device~\cite{Chen2021, Marcks2025} or from one dot at many different locations~\cite{Marcks2025, Volmer2024, Struck2024}. 
We now show that this strategy works well in the deterministic regime but not the disordered regime, even in the limit of impractically large data sets.
(E.g., see \cite{SM}.)

\begin{figure}[t]
\begin{center}
\includegraphics[width=0.48\textwidth]{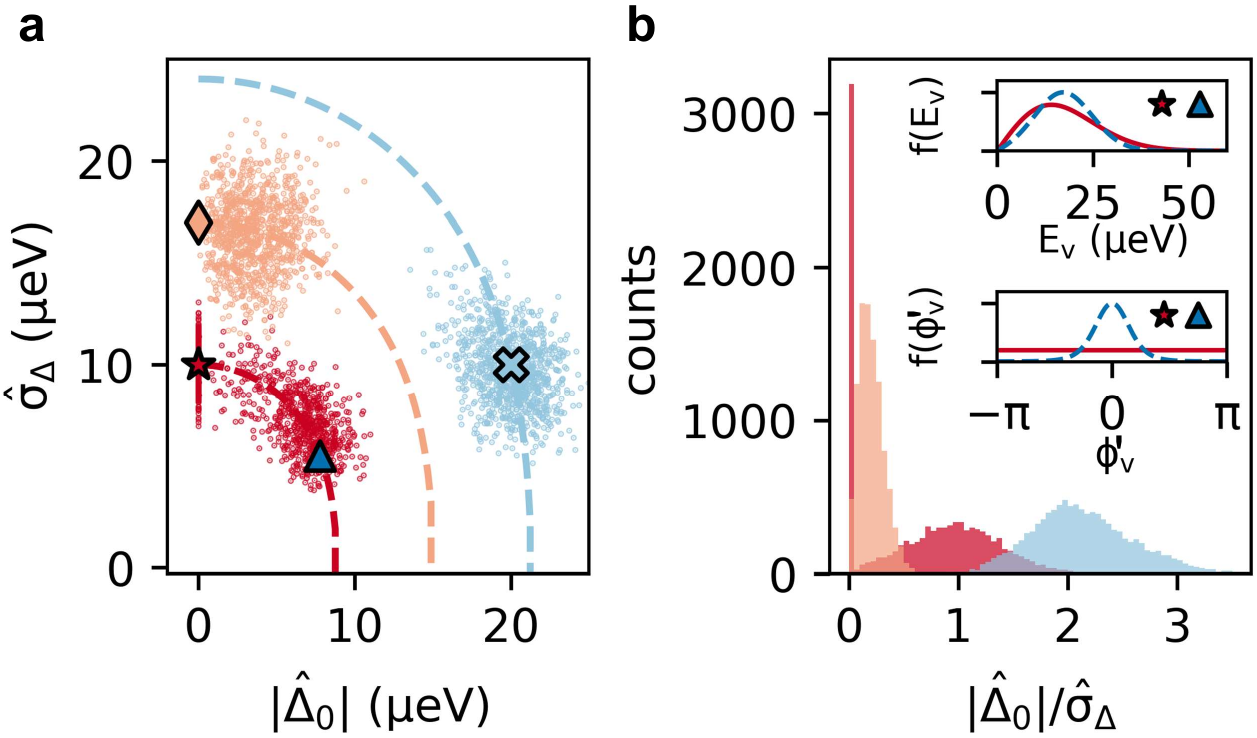}
\end{center}
\vspace{-0.5cm}
\caption{
Statistical characterization of valley splitting distributions.
(a) Valley splittings are generated randomly from the distribution function in Eq.~(\ref{EvDis}) for $10^{4}$ groups of 20 dots.
Each group of 20 is used to reconstruct the original distribution via maximum likelihood estimation, yielding the best-fit parameters $|\hat{\Delta}_0|$ and $\hat{\sigma}_{\Delta}$.
Red and blue data show results obtained for the Rician distribution, Eq.~(\ref {EvDis}), with generating parameters ($|\Delta_0|$, $\sigma_\Delta$)=(0, 10)~$\mu$eV ($\star$ symbol) or (20, 10)~$\mu$eV ($\times$ symbol), respectively.
Salmon data show results for a complex normal distribution, Eq.~(\ref{fDelta}), with generating parameters ($|\Delta_0|$, $\sigma_\Delta$)=(0, 17)~$\mu$eV and $\phi_0$=(arbitrary, because $|\Delta_0|=0$) ($\diamond$ symbol).
Dashed lines show families of generating parameters that yield the same average $\overline E_v$, including the parameters used to generate the data.
Note that the blue triangle ($\triangle$) has the same $\bar{E}_v$ as $\star$, but is not used for generating data.
(b) Distributions of the parameter ratio $|{\hat\Delta}_0|/{\hat\sigma}_\Delta$, taken from (a). 
Note that the red distribution incorrectly indicates deterministic behavior, with the majority of points showing $|{\hat\Delta}_0|/{\hat\sigma}_\Delta>\sqrt{\pi}/2$.
When only $E_v$ data are available, the distributions can be characterized accurately in the deterministic regime (blue data), but not the disordered regime (red data).
When $\phi_v$ data are also available, the distributions can be characterized more accurately (salmon data), meaning the inferred ratio of $|{\hat\Delta}_0|$ and ${\hat\sigma}$ is correctly found to be in the disordered regime ($|{\hat\Delta}_0|/{\hat\sigma} < \sqrt{\pi}/{2}$) in all cases.
Upper inset: $E_v$ distributions for two sets of generating parameters in (a), indicated by the star and triangle ($\triangle$), having the same average $\overline E_v$; for common sample sizes, these distributions are difficult to distinguish.
Lower inset: $\phi_{v}^\prime = \phi_v - \phi_0$ distributions for the same sets of generating parameters; here, these two distributions are easy to distinguish, which explains why valley splitting distributions are easier to characterize when $\phi_v$ information is available.
}
\label{FIG1}
\vspace{-1mm}
\end{figure}

In Fig.~\ref{FIG1}, we generate random samples from Eq.~(\ref{EvDis}), and then attempt to reconstruct the distribution from these data.
We begin by randomly sampling $E_v$ for a statistically independent set of quantum dots, for two cases: $|\Delta_0|/\sigma_{\Delta} = 2$ (deterministic regime) and $|\Delta_0|/\sigma_{\Delta} = 0$ (disordered regime).
We then divide these data into groups of 20, where each group is meant to represent a large but realistic experimental sample.
For each group of 20 dots, we then use maximum likelihood to estimate the best-fit parameters for the Rician distribution, denoted as $|\hat{\Delta}_0|$ and $\hat{\sigma}_{\Delta}$. 
Note that we resort to numerical techniques in solving for $|\hat{\Delta}_0|$ and $\hat{\sigma}_{\Delta}$, as solving for them analytically appears quite difficult, if not impossible (see \cite{SM}).
Results are shown in Fig.~\ref{FIG1}(a), where each data point represents a single group of 20 dots, with blue and red indicating the deterministic and disordered cases, respectively;
the blue $\times$ and red $\star$ show the corresponding generating parameters for the two distributions.
Each color shows data for $10^4$ groups.
Note that different combinations of generating parameters $|\Delta_0|$ and $\sigma_\Delta$ can yield the same average valley splitting $\overline E_v$; the dashed lines in Fig.~\ref{FIG1}(a) indicate three such families, which include the generating parameters for the data shown in the figure.
We see that the red and blue data fall near the red and blue curves, as expected.
The deterministic (blue) data also fall near their corresponding generating parameter (blue $\times$).
However, a majority~\footnote{We also observe an enhancement of counts with $|\hat{\Delta}_0| = 0$, caused by $|\hat{\Delta}_0| = 0$ being an endpoint of the fitting procedure.} of red data points fall far from their generating parameter (red ($\star$)), thus vastly overestimating $|\Delta_0|$.
In fact, for realistically sized data sets, the method typically misclassifies disordered samples as deterministic. 
This dramatic overestimation of the deterministic valley coupling in the disordered regime also occurs when including an impractically larger number of dots per group (see \cite{SM}).

\begin{figure*}[t]
\begin{center}
\includegraphics[width=0.96\textwidth]{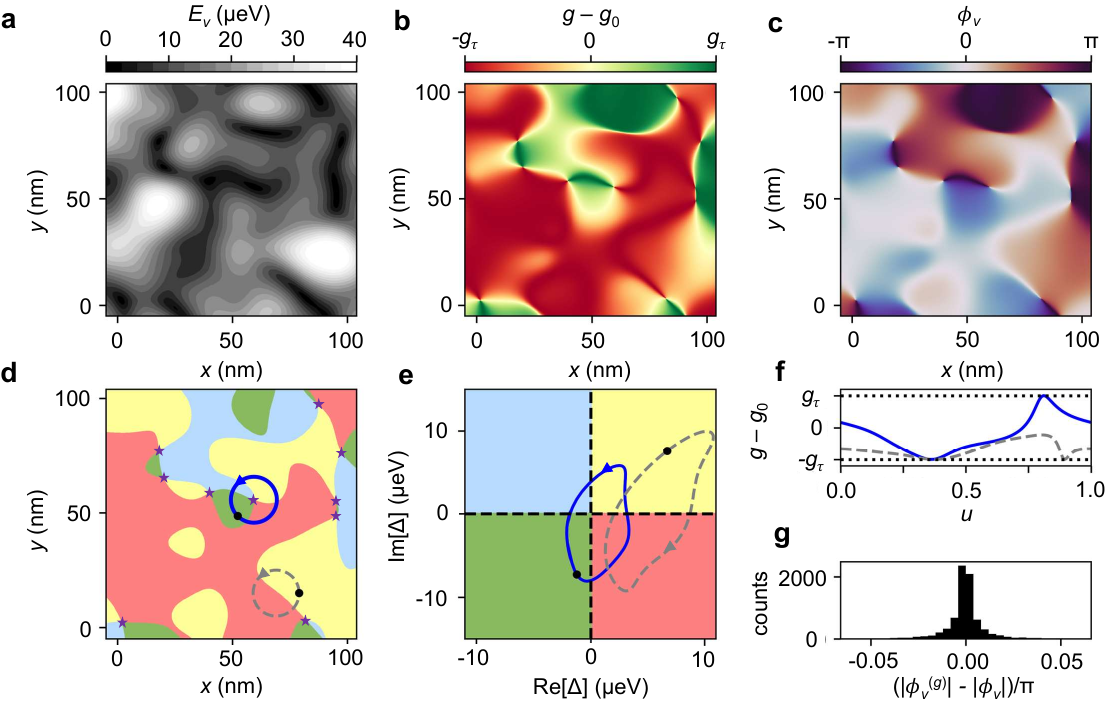}
\end{center}
\vspace{-0.5cm}
\caption{
Probing the valley phase using $g$-factor measurements.
(a)-(c) Three two-dimensional landscapes are computed for a Si/SiGe dot of radius $\ell = \sqrt{\hbar/(m_t \omega_t)} = 14.2~\T{nm}$, for the same alloy disorder realization, with valley coupling parameters $(\Delta_0, \sigma_{\Delta}) = (0,20)~\mu\T{eV}$. 
(a) Valley splitting ($E_v$) landscape. 
(b) $g$-factor landscape. 
(c) Valley phase ($\phi_{v}$) landscape. 
Here, $E_v$ and $g$ are experimentally measurable, while $\phi_v$ can be determined from $g$-factor data by inverting Eq.~(\ref{gFactor}). 
(d) The valley phase in (c) is replotted in a four-color scale, with different colors representing different quadrants of the complex $\Delta$ plane.
Valley vortices (VVs) occur at points where the four colors meet (purple stars), corresponding to points where $E_v = 0$. [Compare with (a).]
We propose to determine the maximum deviations of $g$ from its average value $g_0$, and thus determine the value of $g_\tau$ in Eq.~(\ref{gFactor}), by measuring $g$ around a loop containing a VV. 
(e) The same two paths shown in (d), plotted on the complex $\Delta$ plane.
(f) Fluctuations of the $g$-factor, $g\! - \! g_0$, along the same two paths, as a function of a relative position variable $u$. 
Since the solid blue loop encloses one VV, $g - g_0$ is guaranteed to visit its maximum and minimum values ($\pm g_\tau$) along this path. 
(g) A statistical distribution of valley-phase estimation errors $|\phi^{(g)}_v| \! - \! |\phi_{v}|$, for 5,000 disorder realizations, where $\phi_{v}$ is the actual valley phase obtained for a ground-state wavefunction and $\phi^{(g)}_v$ is the approximate valley phase obtained by inverting Eq.~(\ref{gFactor}) while ignoring the effects of random alloy disorder in $g_\tau$. 
The resulting errors are very small, with a standard deviation of $0.012\pi$. 
}
\label{FIG2}
\vspace{-1mm}
\end{figure*}

On the other hand, simultaneous measurements of $E_v$ \emph{and} $\phi_v$ provide a means of characterizing the valley splitting distribution more accurately.
To see this, we follow the same procedure as before, generating data from a distribution, collecting the data into groups of 20, and using maximum likelihood to reconstruct the original distribution.
However, we now use the $\Delta$ distribution from Eq.~(\ref{fDelta}), rather than the $E_v$ distribution from Eq.~(\ref{EvDis}). 
The results are shown as salmon data points in Fig.~\ref{FIG1}, again for the case of $|\Delta_0|/\sigma_\Delta=0$~\footnote{Note that there is no enhancement of counts at $|\hat{\Delta}_0| = 0$ in this case, because this is no longer an endpoint in the fitting procedure.}; for visual clarity here, we use a different $\sigma_\Delta$ value than the red data.
We see that the second method does a much better job of estimating $|\Delta_0|$ and $\sigma_\Delta$ than before, when only $E_v$ data was used.
Indeed, for the results shown here, no data are misclassified as deterministic.
\red{See \cite{SM} for further fitting results with intermediate values of $|\Delta_0|/\sigma_{\Delta}$.}

The reason for the improved inferential power that comes from knowledge of $\phi_v$ is illustrated in the insets of Fig.~\ref{FIG1}(b).
The top inset shows two $E_v$ distributions having the same average $E_v$, but very different generating parameters $|\Delta_0|$ and $\sigma_\Delta$, corresponding to the red star and blue triangle in Fig.~\ref{FIG1}(a). 
Despite their differences, the distributions have strikingly similar shapes, rendering them nearly indistinguishable.
In contrast, the second inset shows
$\phi_v^\prime\! =\! \phi_v\! -\! \phi_0$ distributions for the same generating parameters, which are easy to distinguish.
(See \cite{SM} for a derivation of the $\phi_v^\prime$ distribution.)


\textbf{Probing the valley phase with $\bm g$-factor measurements.~--~}
Although $\phi_v$ cannot be probed directly, there is a direct relation between $\phi_v$ and the fluctuations of $g$ about its mean value $g_0$, which can be measured accurately in qubit experiments.
We now review this relationship, which was first derived in~\cite{Woods2026a}, and is also summarized in~\cite{SM}.

An effective Hamiltonian can be derived from an effective-mass model by first evaluating matrix elements in a basis labeled with spin, valley, and orbital quantum numbers, followed by projecting onto the lowest-energy orbital within a second-order Schrieffer-Wolff transformation~\cite{Woods2026a}.
This results in a $4 \times 4$ effective Hamiltonian given by
\begin{equation}
    \begin{split}
    H_{\T{eff}} =&~ \Delta \tau_- + \Delta^* \tau_+
    + \frac{\mu_B g_0}{2} {\mathbf B} \cdot \BS{\sigma} \\
    &+ \frac{\mu_B}{2}\left( g_{\tau} \tau_- + g_{\tau}^* \tau_+\right)
    \left(B_x \sigma_y + B_y \sigma_x\right).
\end{split}\label{Heff}
\end{equation}
Here, $\BS{\sigma}$ is a vector of Pauli matrices in spin space and $\tau_{\pm} = \ket{\pm z}\bra{\mp z}$ are raising/lowering operators for the $\pm z$ valley states in the conduction band of Si.
The matrix elements $\Delta$ and $g_\tau$ describe spin-conserving and spin-nonconserving valley couplings, respectively, including perturbative corrections from orbitally excited states up to second order.
The Zeeman term describes an in-plane magnetic field $\mathbf B$, where  $\mu_B$ is the Bohr magneton and $g_0\approx 2$ is the bare Land\'e $g$-factor.
Note that $g_{\tau} \sim 10^{-3}$ \cite{Woods2026a}, so $g_{\tau} \ll g_0$, and we may assume that $g_{\tau} \in \mathbb{R}_+$ without loss of generality~\footnote{More generally, $g_{\tau} \in \mathbb{C}$; however, the zero of the valley phase can be redefined to give $g_{\tau} \in \mathbb{R}_+$, without loss of generality.}.
We stress that the effective-mass model used as the starting point in deriving the effective Hamiltonian yields spin-orbit coefficients, g-factors, and valley splittings that are all in good agreement with accurate $sp^3d^5s^*$ tight-binding models~\cite{Woods2023a,Losert2023}, and hence we expect the effective Hamiltonian in Eq. (\ref{Heff}) to capture all of the relevant physics.

To compute the $g$-factor, we diagonalize the $\Delta$ terms of Eq.~(\ref{Heff}) in the limit $B\rightarrow 0$ and project onto the valley ground state.
We diagonalize the resulting two-dimensional spin Hamiltonian to obtain the ground and excited energies $E_0$ and $E_1$.
Defining $g \!=\! (E_1 \! - \! E_0)/(\mu_B B)$ when $B\rightarrow 0$, we obtain
\begin{equation}
    g = g_0 - g_{\tau} \cos\phi_v \sin(2\theta_B), 
    \label{gFactor}
\end{equation}
where $\theta_B$ is the angle of $\mathbf B$ with respect to the $[100]$ crystallographic axis, i.e. $\BS{B} = B (\cos\theta_B\hat{x} + \sin\theta_B \hat{y})$.
For experimental protocols, we may choose $\theta_B = \pi/4$ (i.e. $\BS{B}$ along the $[110]$ crystallographic axis) to maximize the size of $g$-factor fluctuations.
\red{Importantly, recent shuttling experiments \cite{Volmer2026} provide strong evidence for the validity of Eq.~\eqref{gFactor}.} 

In Eqs.~(\ref{Heff}) and (\ref{gFactor}), we note that the amplitude and phase of the valley coupling vary in the plane of the quantum well due to alloy disorder, $\Delta = \Delta(x,y)$, yielding valley splitting landscapes like the one shown in Fig.~\ref{FIG2}(a), where $(x,y)$ denotes the center of a quantum dot. 
The dependence of $g$ on $\phi_v$ then gives rise to $g$-factor landscapes like the one shown in Fig.~\ref{FIG2}(b).
See Ref. \cite{Woods2026a} for a description of the method used for generating $E_v$ and $\phi_v$ landscapes.
The central idea of this work is to invert Eq.~(\ref{gFactor}) to extract $\phi_v$, as shown in Fig.~\ref{FIG2}(c), through measurements of $g$. 
There are two potential challenges for implementing this procedure, which we now address.

\textbf{Using valley vortices to determine $\bm g_{\bm \tau}$.~--~}
The first challenge with inverting Eq.~(\ref{gFactor}) is that $g_\tau$ is not known, a priori.
An inefficient solution would be to map out $g$ over a region large enough to include its extreme values with high probability.
As a more efficient alternative, we propose to determine $g_\tau$ \emph{with certainty} by measuring $g$ around a loop enclosing a single valley vortex (VV), which we define as a point where the valley splitting vanishes: $E_{v}\! =2|\Delta| = 0$.
VVs can be seen in valley-splitting landscapes, like the one shown in Fig.~\ref{FIG2}(a); however, they are easier to visualize in a map like Fig.~\ref{FIG2}(d), where the four colors denote the four quadrants of the complex plane of $\Delta$ values; here, the VVs occur at points where the four colors meet. 
In Fig.~\ref{FIG2}(d), there are ten such VVs, denoted by purple stars. 

To illustrate our VV proposal, we consider the two loops indicated in the spatial map of Fig.~\ref{FIG2}(d).
Here, the dashed-gray and solid-blue loops enclose zero or one VV, respectively. 
In Fig.~\ref{FIG2}(e), we show the same paths plotted in the complex $\Delta$ plane, where it is clear that they are \textit{topologically} distinct, since the solid-blue loop winds around the origin once, while the dashed-gray loop does not. 
The presence of a non-zero winding number is key to the procedure, as it causes $\phi_v$ to wind by $\pm 2\pi$, ensuring that $g$ achieves its extreme values of $g_0 \pm g_{\tau}$ somewhere along the path.
This is illustrated in the $g - g_0$ plot shown in Fig.~\ref{FIG2}(f), where the solid-blue loop attains both extrema, while the dashed-gray loop attains only its minimum. 

This scheme is efficient for finding $g_\tau$.
However, to implement it, we must first locate a VV. 
A brute force scan offers an obvious, but inefficient solution.
As a more efficient alternative, we propose following a curve of constant $g$~\footnote{In Fig.~\ref{FIG2}(b), such curves are obtained by starting at a random point and following a path that remains at the same color as the starting point. A curve of constant $g$ could be found by measuring the gradient of $g$ using $\nabla g(x,y) \approx \frac{1}{a}[g(x+a,y) - g(x,y)]\hat{x} + \frac{1}{a}[g(x,y+a) - g(x,y)]\hat{y}$, where $a$ is a small displacement, and then moving the electron perpendicular to $\nabla g$.}.
There are only two outcomes for this procedure:
(1) the curve passes through a VV (since the full range of $\phi_v$ values are present here), or
(2) it forms a closed loop without passing through a VV.
The first case is often successful; however, if not, one should try again with a new starting point, and repeat until a VV is located.
We note that it should not be difficult to locate VVs in the disordered regime, since their density is fairly high.
In~\cite{SM}, we derive this density, obtaining
\begin{equation}
\rho_\text{VV} = (0.184/ \ell^2) e^{-|\Delta_0|^2/ \sigma_{\Delta}^2}, \label{rhovvMain}
\end{equation}
where $\ell = \sqrt{\hbar/(m_t \omega_t)}$ is the quantum dot radius.
Indeed, for the $100 \times 100~\T{nm}^2$ region considered in Fig.~\ref{FIG2}, Eq.~(\ref{rhovvMain}) predicts the occurrence of 9.2 VV, on average, while 10 VV are observed here.
Note that
the VV density scales with dot size, such that any dot should find the same number of VV in its vicinity (on average), regardless of its size.

\textbf{Higher-order fluctuations in $g$.~--}~
The second challenge with inverting Eq. (\ref{gFactor}) is that it neglects higher-order effects of alloy disorder. 
For example, in Eq.~(\ref{Heff}), the $\Delta$ term includes disorder directly in its random component.
However, the leading corrections to the $g$-factor from $g_\tau$ are deterministic: they only depend on disorder at third order in the perturbation theory.
Fortunately, this means that disorder effects are weak in typical scenarios, as illustrated in Fig.~\ref{FIG2}(g).
Here, we calculate $g$ for a physically realistic Si/SiGe quantum well, as described in~\cite{SM}.
We then compare the valley phase computed via two different methods.
In the first method, we generate the valley coupling landscape for one instance of alloy disorder and directly extract the absolute value of the valley phase $|\phi_v|$ (see Eq. (S.54) of \cite{SM}).
Second, we use the same alloy disorder to compute $g \!=\! (E_1 \! - \! E_0)/(\mu_B B)$ in the limit $B\rightarrow 0$ by directly solving for the eigenvalues $E_0$ and $E_1$, as described in \cite{SM}. 
Note that the obtained eigenvalues contain the information from the higher-order perturbations neglected in Eq. (\ref{Heff}).
We then invert Eq.~(\ref{gFactor}), using the $g_\tau$ value computed in the absence of disorder, obtaining the approximate result $|\phi^{(g)}_v|$.
The difference $|\phi^{(g)}_v| - |\phi_{v}|$ represents the error induced by neglecting higher-order corrections from alloy disorder.
A distribution of $|\phi^{(g)}_v| - |\phi_{v}|$ values for 5,000 different disorder realizations is shown in Fig.~\ref{FIG2}(g).
The errors are seen to be small, with a standard deviation of $0.012\pi$ that represents the uncertainty of the $\phi_v$ extraction procedure.
Further discussion of this distribution is given in~\cite{SM}.

\textbf{Conclusion.~--}~
We have shown that sampling a distribution of $E_v$ values, with no other information, yields very poor estimates of the valley coupling parameters $|\Delta_0|$ and $\sigma_{\Delta}$, except in rare cases where $|\Delta_0|\gtrsim \sigma_{\Delta}$.
However, we have also shown that this situation is greatly improved when valley phase $\phi_{v}$ information is available. 
Motivated by this fact, we proposed to extract $\phi_{v}$ from $g$-factor measurements across the quantum well.
A key calibration step for this procedure involves measuring the spatial variations of $g$ around a path enclosing a valley vortex, because this reveals the full range of $g$ values.
Since characterizing $|\Delta_0|$ and $\sigma_{\Delta}$ in Si/SiGe quantum wells is a key step for enabling spin-based quantum processors at scale, we expect the current proposal to provide an important characterization tool for near-term experiments. 
Importantly, the proposal should be possible to carry out with current experimental shuttling devices, such as the device in Ref. \cite{Volmer2024}, which provide a spatial measurement resolution on order of a nm and $E_v$ and $g$-factor measurement error bars that are less than $1~\mu\T{eV}$ and $10^{-5}$, respectively.

\textbf{Acknowledgments.~--}~This research was sponsored in part by the Army Research Office (ARO) under Awards No.\ W911NF-17-1-0274, W911NF-22-1-0090, and W911NF-23-1-0110. The views, conclusions, and recommendations contained in this document are those of the authors and are not necessarily endorsed nor should they be interpreted as representing the official policies, either expressed or implied, of the ARO or the U.S. Government. The U.S. Government is authorized to reproduce and distribute reprints for Government purposes notwithstanding any copyright notation herein.


%

\clearpage

\renewcommand\thefigure{S.\arabic{figure}}
\renewcommand\thesection{S.\arabic{section}}
\renewcommand\theequation{S.\arabic{equation}}
\setcounter{figure}{0}
\setcounter{equation}{0}
\setcounter{section}{0}
\onecolumngrid

\section*{Supplementary Materials}

\section{Maximum likelihood equations for $|\hat{\Delta}_0|$ and $\hat{\sigma}_{\Delta}$}
In the main text, we employed maximum likelihood estimation to determine the best fit parameters $|\hat{\Delta}_0|$ and $\hat{\sigma}_{\Delta}$ given a finite sample of $E_v$ values. There we commented that we have resorted to numerical techniques in solving for $|\hat{\Delta}_0|$ and $\hat{\sigma}_{\Delta}$, as solving for them analytically appears quite difficult, if not impossible. In this Supplementary Materials section, we illustrate this difficulty by deriving the maximum likelihood equations that $|\hat{\Delta}_0|$ and $\hat{\sigma}_{\Delta}$ satisfy.

Our problem is finding the maximum likelihood parameters ($|\hat{\Delta}_0|,\hat{\sigma}_{\Delta}$) given a group of valley splitting data $E =\{E_{v,1},E_{v,2},\dots,E_{v,N}\}$, where $E_{v,i}$ is the valley splitting of dot $i$ and $N$ is the number of dots in the data set. Finding the maximum likelihood parameters ($|\hat{\Delta}_0|,\hat{\sigma}_{\Delta}$) amounts to finding ($|\Delta_0|,\sigma_{\Delta}$) such that $f_{E}(E;|\Delta_0|,\sigma_{\Delta})$ is maximized, where $f_{E}(E;|\Delta_0|,\sigma_{\Delta})$ is the conditional probability density of obtaining valley splitting data $E = \{E_{v,1},E_{v,2},\dots,E_{v,N}\}$ given the parameters $(|\Delta_0|,\sigma_{\Delta})$. Assuming the dots are statistically independent (due to being sufficiently separated), the conditional probability factorizes as
\begin{equation}
    f_{E}(E;|\Delta_0|,\sigma_{\Delta}) = \prod_{i=1}^{N} f_{E_{v}}(E_{v,i};|\Delta_0|,\sigma_{\Delta}), 
\end{equation}
where $f_{E_{v}}(E_{v};|\Delta_0|,\sigma_{\Delta})$ is the valley splitting probability density function for a single dot given in Eq. (2) of the main text. We can transform the product into a sum by maximizing the natural log of $f_{E}$, $\mathcal{L}(E;|\Delta_0|,\sigma_{\Delta}) \equiv \ln\left[f_{E}(E;|\Delta_0|,\sigma_{\Delta})\right]$, which is found to be
\begin{equation}
    \mathcal{L}(E;|\Delta_0|,\sigma_{\Delta}) =
     \sum_{i=1}^{N} \ln\left(\frac{E_{v,i}}{2}\right) - 
    2N\ln(\sigma_{\Delta}) + 
    \sum_{i=1}^{N}
    \left(
    - \frac{E_{v,i}^2 + 4|\Delta_0|^2}{4\sigma_{\Delta}^2}
    + \ln\left[
    I_0\left(\frac{E_{v,i}|\Delta_0|}{\sigma_{\Delta}}^2\right)
    \right]
    \right).
\end{equation}
Possible points $(|\Delta_0|,\sigma_{\Delta})$ that maximize $\mathcal{L}$ satisfy
\begin{align}
    \frac{\partial \mathcal{L}}{\partial\sigma_{\Delta}} =& -\frac{2N}{\sigma_{\Delta}}
    + \sum_{i=1}^N
    \left(
    \frac{E_{v,i}^2 + 4|\Delta_0|^2}{2\sigma_{\Delta}^3}
    - \frac{E_{v,i}|\Delta_0|}{2\sigma_{\Delta}^3}
    \frac{I_1\left(\frac{E_{v,i}|\Delta_0|}{\sigma_{\Delta}^2}\right)}
    {I_0\left(\frac{E_{v,i}|\Delta_0|}{\sigma_{\Delta}^2}\right)}
    \right) = 0, \\
    \frac{\partial \mathcal{L}}{\partial|\Delta_0|} =&
    -\frac{2N|\Delta_0|}{\sigma_{\Delta}^2} + 
    + \sum_{i=1}^N
    \frac{E_{v,i}}{\sigma_{\Delta}^2}
    \frac{I_1\left(\frac{E_{v,i}|\Delta_0|}{\sigma_{\Delta}^2}\right)}
    {I_0\left(\frac{E_{v,i}|\Delta_0|}{\sigma_{\Delta}^2}\right)} = 0.
\end{align}
Finding all the solutions ($|\Delta_0|,\sigma_\Delta$) analytically that satisfy these complicated maximum likelihood equations, and then finding the global maximum among them seems quite difficult to us, if not impossible. Furthermore, even if we could find such a solution analytically, it would be a complicated $N$-dimensional function, as it depends on $E_{v,1}, E_{v,2},\dots, E_{v,N}$.

In contrast, a numerical solution to the maximum likelihood equations can be readily found by standard gradient-descent algorithms. Furthermore, we can understand the distribution of $(|\hat{\Delta}_0|,\hat{\sigma}_{\Delta})$ by solving this maximization problem for many realizations of $E = \{E_{v,1},E_{v,2},\dots,E_{v,N}\}$, as we have done to obtain the results shown in Fig. 1 of the main text.

\begin{figure}[t]
\begin{center}
\includegraphics[width=0.48\textwidth]{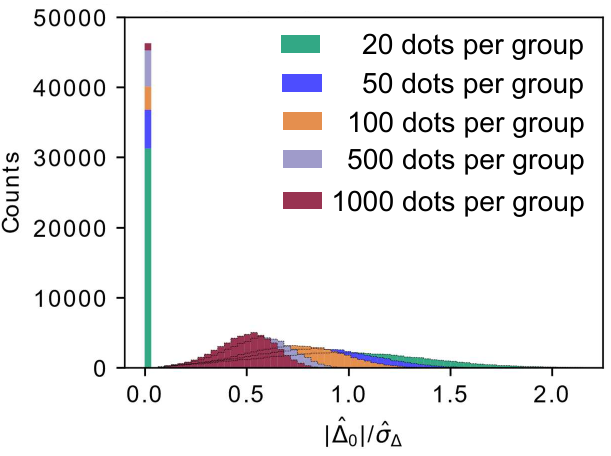}
\end{center}
\vspace{-0.5cm}
\caption{Histogram of the distribution of best-fit estimates for the ratio $|\hat{\Delta}_0|/\hat{\sigma}_{\Delta}$, obtained for 100,000 groups of dots with generating parameters ($|\Delta_0|$, $\sigma_\Delta$) = (0, 10)~$\mu$eV. The five colors (green, blue, orange, purple, red) correspond to different numbers $N$ of dots in the group, with $N = 20, 50, 100, 500$, and 1000. Note that $N = 20$ was used for the results in Fig.~1 of the main text. We see that even for the case of $N = 1000$ dots per group, which is well beyond the size of modern quantum dot arrays, a significant, non-zero deterministic component $|\hat{\Delta}_0|$ is inferred in over half of the cases. Note that the ordering of histograms are inverted for the bin near $|\hat{\Delta}_0|/\hat{\sigma}_{\Delta} = 0$ such that the data for each $N$ is visible.}
\label{FIGS0}
\vspace{-1mm}
\end{figure}

\section{Characterizing the valley splitting distribution with a larger number of dots}
In Fig.~1 of the main text, for the red and blue data, we performed maximum likelihood estimation using groups of $N = 20$ dots to determine the best-fit parameters $|\hat{\Delta}_0|$ and $\hat{\sigma}_{\Delta}$ describing those valley splitting distributions. 
From those results, we concluded that $|\Delta_0|$ and $\sigma_{\Delta}$ can be determined with reasonable accuracy from $E_v$ data alone, if they are in the deterministically enhanced regime, but not when they are in the disordered regime. In this Supplementary Materials section, we show that this conclusion continues to hold true, even when an impractically large number of dots are included in the analysis.
To do so, we perform the same maximum likelihood estimation analysis as the main text, with generating parameters ($|\Delta_0|$, $\sigma_\Delta$) = (0, 10)~$\mu$eV corresponding to the completely disordered regime.
Here, however, we vary the number of dots $N$ in each group.
The resulting best-fit ratios $|\hat{\Delta}_0|/\hat{\sigma}_\Delta$ are shown in Fig.~\ref{FIGS0} for 100,000 groups of size $N = 20$ (green), $50$ (blue), $100$ (orange), $500$ (purple), and $1000$ (red) dots. 
We obtain results for $N=20$ that are in agreement with the red data of Fig.~1 of the main text, as expected.
For larger group sizes, the best-fit results slowly approach the correct answer ($\Delta_0/\sigma_\Delta=0$).
However, even in the case of $N = 1000$ dots per group, which is well beyond the size of state-of-the-art quantum dot arrays, a ratio $|\hat{\Delta}_0|/\hat{\sigma}_{\Delta}$ significantly greater than $0$ is incorrectly inferred in over half of the cases.

\section{Characterizing the valley splitting distribution for finite $\Delta_0$}

\begin{figure}[t]
\begin{center}
\includegraphics[width=0.9\textwidth]{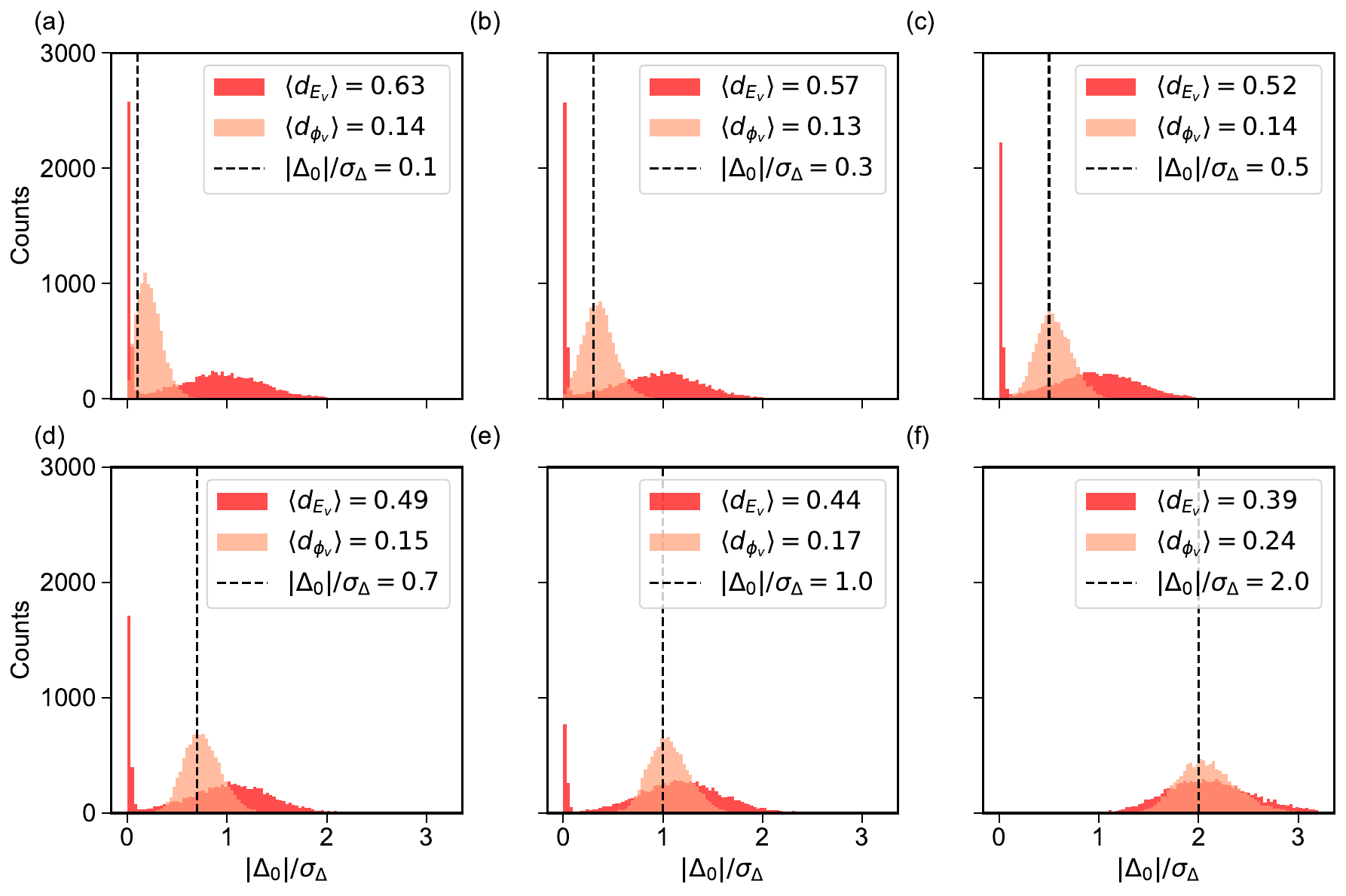}
\end{center}
\vspace{-0.5cm}
\caption{\red{Histograms of the best-fit estimates for the ratio $|\hat{\Delta}_0|/\hat{\sigma}_{\Delta}$ for $|\Delta_0|/\sigma_{\Delta} = \{0.1,0.3,0.5,0.7,1.0,2.0\}$. Red data includes just $E_{v}$ data, while the salmon data includes both $E_{v}$ and $\phi_{v}$ data for each dot. Each panel shows results for $10,000$ groups of $N = 20$ dots. The dashed lines correspond to the $|\Delta_0|/\sigma_{\Delta}$ values used to generate the data for each panel. 
The average error distances $\left<d_{E_{v}}\right>$ and $\left< d_{\phi_{v}}\right>$ defined in Eq.~(\ref{errorDis}) are shown for each $|\Delta_{0}|/\sigma_{\Delta}$ value. In all cases, we see that knowledge of $\phi_{v}$ significantly decreases the average inference error.}
} 
\label{FIGSFiniteDelta0}
\vspace{-1mm}
\end{figure}

\red{In Fig. 1 of the main text, we performed maximum likelihood for valley coupling ratios of $|\Delta_0|/\sigma_\Delta = 0,2$. In this supplementary Materials section, we extend this analysis to intermediate and finite values of $|\Delta_0|/\sigma_\Delta$. The results are shown in Fig.~\ref{FIGSFiniteDelta0}(a-f) for $|\Delta_0|/\sigma_{\Delta} = \{0.1,0.3,0.5,0.7,1.0,2.0\}$. Here, the red data uses only $E_{v}$ data when performing the maximum likelihood estimation, like the blue and red data in Fig.~1 of the main text. In contrast, the salmon data uses both $E_{v}$ and $\phi_v$ data when performing the maximum likelihood estimation, like the salmon data in Fig.~1 of the main text. As is the case for Fig.~1 of the main text, the maximum likelihood estimation results in Fig.~\ref{FIGSFiniteDelta0} use $N = 20$ dots per group. Each panel shows results for $10,000$ groups of $N = 20$ dots. The dashed lines correspond to the $|\Delta_0|/\sigma_{\Delta}$ values used to generate the data for each panel.}

\red{For Fig.~\ref{FIGSFiniteDelta0}(a-e), corresponding to $|\Delta_0|/\sigma_{\Delta} = \{0.1,0.3,0.5,0.7,1.0\}$, we see that the distribution of $|\hat{\Delta}_0|/\hat{\sigma}_{\Delta}$ barely changes when only $E_{v}$ data is known. In contrast, the distribution of $|\hat{\Delta}_0|/\hat{\sigma}_{\Delta}$ significantly changes with $|\Delta_{0}|/\sigma_{\Delta}$ when both $E_{v}$ and $\phi_{v}$ are known. Indeed, except for $|\Delta_{0}|/\sigma_{\Delta} = 0.1$, the mode of the $|\hat{\Delta}_0|/\hat{\sigma}_{\Delta}$ distribution is found at the generating $|\Delta_{0}|/\sigma_{\Delta}$ value when both $E_{v}$ and $\phi_{v}$ are known. In contrast, when only $E_{v}$ data is known, the mode of the $|\hat{\Delta}_0|/\hat{\sigma}_{\Delta}$ distribution is found at the generating $|\Delta_{0}|/\sigma_{\Delta}$ value only in the deterministic regime of $|\Delta_0|/\sigma_{\Delta} = 2$ shown in Fig.~\ref{FIGSFiniteDelta0}(f).
}

\red{
To quantify the performance of the two inference methods, we define the error distance 
\begin{equation}
    d = 
    \left|
    \frac{\hat{\Delta}_0}{\hat{\sigma}_{\Delta}} -
    \frac{{\Delta}_0}{{\sigma}_{\Delta}}
    \right|. \label{errorDis}
\end{equation}
The average error distances are shown in each panel of Fig.~\ref{FIGSFiniteDelta0}. Here, $\left<d_{E_{v}}\right>$ 
and $\left<d_{\phi_{v}}\right>$ are the average error distance when using just $E_{v}$ data and both $E_{v}$ and $\phi_{v}$ data, respectively.
In all cases, we see that inference with knowledge of $\phi_{v}$ significantly improves the average inference error distance.
Furthermore, we see that difference in the average error distance is larger for decreasing $|\Delta|_0/\sigma_{\Delta}$, highlighting the importance of $\phi_{v}$ information in the disordered and intermediate $|\Delta_0|/\sigma_{\Delta}$ regimes.
}

\section{Distribution of complex inter-valley coupling $\Delta$}
To obtain the salmon-color data set in Fig.~1 of the main text, we randomly sample $\Delta$ in groups of 20 dots and perform maximum likelihood estimation of the best-fit parameters $|\hat{\Delta}_0|$ and $\hat{\sigma}_{\Delta}$. 
In this Supplementary Materials section, we provide details regarding the complex distribution of $\Delta$. 

Recall that we can decompose the inter-valley coupling as $\Delta = \Delta_0 + \Delta_{\delta}$, where $\Delta_0$ and $\Delta_{\delta}$ are the deterministic and random components, respectively.
Here, $\Delta_0 = |\Delta_0|e^{i\phi_0}$. The random component $\Delta_\delta$ follows a complex Gaussian distribution with uncorrelated real and imaginary components, both having a variance of $\sigma_\Delta^2/2$ and zero mean \cite{Losert2023}.
It follows that the real ($\Delta_r$) and imaginary ($\Delta_i$) components of $\Delta$ both have normal distributions given by 
\begin{align}
    f(\Delta_r ; |\Delta_0|,\phi_0, \sigma_\Delta) &= \frac{1}{\sqrt{\pi \sigma_{\Delta}^2}}
    e^{-
    \left(\Delta_r - |\Delta_0|\cos\phi_0\right)^2
    /\sigma_{\Delta}^2
    }, \label{fReDelta} \\ 
     f(\Delta_i ; |\Delta_0|,\phi_0, \sigma_\Delta) &= \frac{1}{\sqrt{\pi \sigma_{\Delta}^2}}
    e^{-
    \left(\Delta_i - |\Delta_0|\sin\phi_0\right)^2
    /\sigma_{\Delta}^2
    }. \label{fImDelta}
\end{align}
These distributions are both characterized by the variables $|\Delta_0|,\phi_0$, and $\sigma_\Delta$. 
Our maximum likelihood estimation therefore also includes a best-fit parameter $\hat{\phi}_0$, in addition to the best-fit parameters $|\hat{\Delta}_0|$ and $\hat{\sigma}_{\Delta}$ use for the $E_v$ distribution. 
$\phi_0$ has no impact on the valley splitting $E_v$ distribution, however, so we do not report it in Fig.~1 of the main text. 
Moreover, one can always remove a non-zero $\phi_0$ by a global redefinition of the valley degrees of freedom.

\section{Valley phase probability distribution derivation}
In Fig.~1 of the main text, we plot the distribution of the relative valley phase $\phi_{v}^\prime = \phi_v - \phi_0$, where $\Delta_0 = |\Delta_0|e^{i\phi_0}$.
In this Supplementary Materials section, we derive this distribution.

We first define a ``rotated'' inter-valley coupling $\Delta^\prime = \Delta e^{-i\phi_0} = |\Delta |e^{i\phi_v^\prime}$, such that the rotated deterministic component is real and nonnegative, $\Delta_0^\prime = |\Delta_0|$.
The distribution functions of the real and imaginary components of $\Delta^\prime$ are then given by
\begin{align}
    f(\Delta^\prime_r ; |\Delta_0|, \sigma_\Delta) &= \frac{1}{\sqrt{\pi \sigma_{\Delta}^2}}
    e^{-
    \left((\Delta^\prime_r - |\Delta_0|\right)^2
    /\sigma_{\Delta}^2
    }, \label{fReDelta2} \\ 
    f(\Delta^\prime_i ; \sigma_\Delta) &= \frac{1}{\sqrt{\pi \sigma_{\Delta}^2}}
    e^{-
    (\Delta^\prime_i)^2
    /\sigma_{\Delta}^2
    }. \label{fImDelta2}
\end{align}
We further define $\phi_v^\prime$ to be in the interval $(-\pi,\pi]$.
The cumulative distribution function of $\phi_v^\prime$ is then given by
\begin{equation}
    F(\phi^\prime_v ; |\Delta_0|, \sigma_\Delta)= 
    \int_{-\infty}^\infty \int_{-\infty}^\infty
    f(\Delta^\prime_r ; |\Delta_0|, \sigma_\Delta) 
    f(\Delta^\prime_i ; \sigma_\Delta)
    \Theta(\phi_{v}^\prime - \T{Arg}(\Delta^\prime_r + i \Delta^\prime_i))
    \, d\Delta^\prime_r ~d\Delta^\prime_i, \label{CumphiV1}
\end{equation}
where $\Theta(x)$ is the Heaviside step function, and $\T{Arg}(x)$ is the argument function restricted to its principle value in the interval $(-\pi,\pi]$.
The integral in Eq.~(\ref{CumphiV1}) is easier to handle in polar coordinates, where we find
\begin{align}
    F(\phi^\prime_v ; |\Delta_0|, \sigma_\Delta) &= 
    \frac{1}{\pi \sigma_{\Delta}^2}
    e^{-|\Delta_0|^2/\sigma_\Delta^2}
    \int_{-\pi}^{\pi} 
    \int_{0}^{\infty}
    e^{
    \frac{-r^2 + 2|\Delta_0|r\cos \phi} 
    {\sigma_{\Delta}^2}
    }
    \Theta(\phi_v^\prime - \phi)
    r 
    \, dr d\phi \\
    &=
    \frac{1}{\pi \sigma_{\Delta}^2}
    e^{-|\Delta_0|^2/\sigma_\Delta^2}
    \int_{-\pi}^{\phi_{v}^\prime} 
    \int_{0}^{\infty}
    e^{
    \frac{-r^2 + 2|\Delta_0|r\cos \phi} 
    {\sigma_{\Delta}^2}
    }
    r 
    \, dr d\phi.
    \label{CumphiV2}
\end{align}
The corresponding probability density function, plotted in the lower inset of Fig.~1 in the main text, is then given by
\begin{align}
    f(\phi^\prime_v ; |\Delta_0|, \sigma_\Delta)  = 
    \frac{\partial F({\phi_v^\prime})}
    {\partial \phi_v^\prime} =&
     \frac{1}{\pi \sigma_{\Delta}^2}
    e^{-|\Delta_0|^2/\sigma_\Delta^2} 
    \int_{0}^{\infty}
    e^{
    \frac{-r^2 + 2|\Delta_0|r\cos \phi_v^\prime} 
    {\sigma_{\Delta}^2}
    }
    r 
    \, dr \\
    =& 
    \frac{1}{2\pi}
    e^{-|\Delta_0|^2/\sigma_\Delta^2}
    +
    \frac{|\Delta_0| \cos \phi_v^\prime}
    {2\sqrt{\pi}\sigma_\Delta}
    e^{-|\Delta_0|^2 \sin^2 \phi_v^\prime/\sigma_\Delta^2}
    \left[1
    + \erf\left(\frac{|\Delta_0| \cos \phi_v^\prime}{\sigma_{\Delta}}\right)
    \right], \label{fRelphivSup}
\end{align}
where $\erf(x)$ is the error function.
In terms of the original valley phase variable $\phi_v=\phi_v^\prime+\phi_0$, the distribution function is given by
\begin{equation}
     f(\phi_v ; |\Delta_0|, \sigma_\Delta)  =
    \frac{1}{2\pi}
    e^{-|\Delta_0|^2/\sigma_\Delta^2}
    +
    \frac{|\Delta_0| \cos (\phi_v - \phi_0)}
    {2\sqrt{\pi}\sigma_\Delta}
    e^{-|\Delta_0|^2 \sin^2(\phi_v - \phi_0)/\sigma_\Delta^2}
    \left[1
    + \erf\left(\frac{|\Delta_0| \cos (\phi_v - \phi_0)}{\sigma_{\Delta}}\right)
    \right]. \label{fphivSup}
\end{equation}

\section{Density of valley vortices}

In this Supplementary Materials section, we analytically determine the density $\rho_\text{VV}$ of valley vortices (VVs) as a function of the parameter ratio $|\Delta_0|/\sigma_{\Delta}$. 
This quantity is of interest, because our protocol for extracting the valley phase $\phi_v$ relies on a calibration step in which the $g$-factor is measured in a loop enclosing a VV, so it is helpful to know their distribution: the protocol is only practical if VVs can be found within a moderately small area. 
Below, we show that the density is relatively high.

\begin{figure}[t]
\begin{center}
\includegraphics[width=0.48\textwidth]{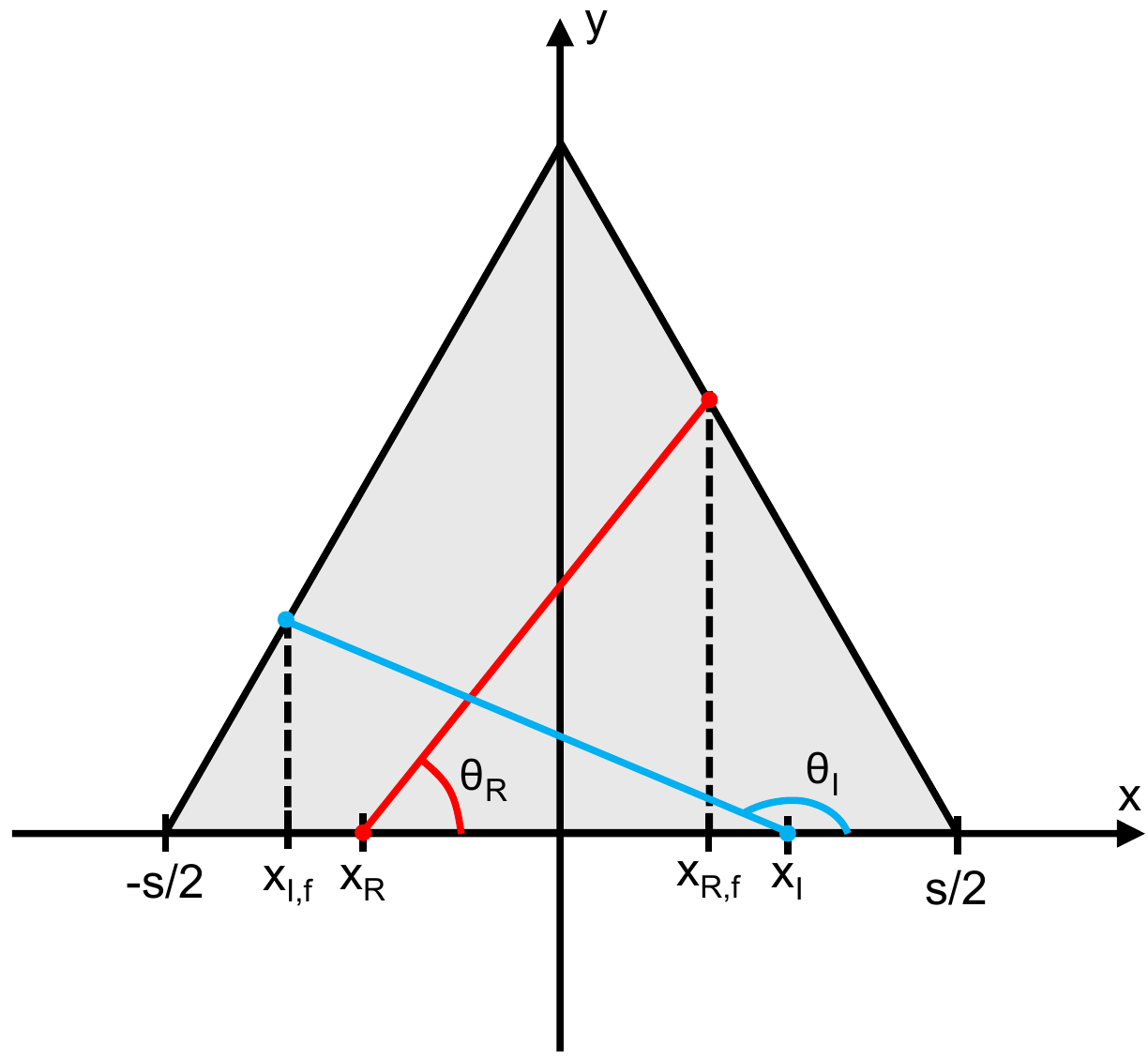}
\end{center}
\vspace{-0.5cm}
\caption{Equilateral triangle of side-length $s$, used to calculate the density of valley vortices. 
Red and blue lines correspond to $\Re[\Delta] = 0$ and $\Im[\Delta] = 0$, respectively. See text in sub-problem 3 for description of the parameters $\{\theta_R,\theta_I,x_R,x_I,x_{R,f},x_{I,f}\}$.}
\label{FIGS1}
\vspace{-1mm}
\end{figure}

To begin, we again decompose the inter-valley coupling as $\Delta(x,y) = \Delta_0 + \Delta_{\delta}(x,y)$, where $\Delta_0$ and $\Delta_{\delta}$ are the deterministic and random components, respectively.
Without loss of generality, we assume $\Delta_0 \in \mathbb{R}_{0}^{+}$. 
Let us also decompose $\Delta_{\delta}$ into real and imaginary components, $\Delta_{\delta}(x,y) = R(x,y) + iI(x,y)$.
The statistics of $\Delta$ depends on the spatial correlations of $R$ and $I$ across the quantum well. 
These correlations, first discussed in Ref.~\cite{Losert2023}, are given by
\begin{align}
    \left< R(\BS{r}_1) R(\BS{r}_2) \right> =& 
    \left< I(\BS{r}_1) I(\BS{r}_2) \right> =
    \frac{\sigma_{\Delta}^2}{2}
    e^{-|\BS{r}_1 - \BS{r}_2|^2/(2\ell_t^2)}, \\
    \left< R(\BS{r}_1) I(\BS{r}_2) \right> =& ~0.
\end{align}
Here, we assume a circular dot with transverse radius $\ell_t = \sqrt{\hbar/(m_t \omega_t)}$.

To determine the density of VVs, we consider an equilateral triangle of side-length $s$, as shown in Fig.~\ref{FIGS1}.
Let us denote the probability of finding $n$ VVs in this triangle as $P_{n}^{\triangle}(s)$.
It should become less likely to observe more than one vortex inside the triangle when triangles are small; thus,
\begin{equation}
    \lim_{s \rightarrow 0} \frac{P_{n}^\triangle(s)}{P_{1}^\triangle(s)} = 0,
\end{equation}
for $n \geq 2$, and the vortex density is determined by $P_1^\triangle$, such that
\begin{equation}
    \rho_\text{VV} = 
    \lim_{s \rightarrow 0} \frac{4 P_1^\triangle(s)}{\sqrt{3} s^2}, \label{rhoVV1}
\end{equation}
where $\sqrt{3}s^2/4$ is the area of the equilateral triangle. 
Therefore, our problem reduces to finding $P_1^\triangle(s)$ to order $\mathcal{O}(s^2)$.
Note that the necessary condition for a point $(x,y)$ to be a valley vortex is that $\Re[\Delta(x,y)] = \Im[\Delta(x,y)] = 0$.

To compute $P_1^\triangle(s)$, we break the problem down into three sub-problems. 
The first sub-problem is to determine the probability that a line segment of length $s$ contains a point $(x,y)$ for which $\Re[\Delta(x,y)] = 0$. 
The second sub-problem is to determine the probability that lines defined by $\Re[\Delta] = 0$ or $\Im[\Delta] = 0$ cross through an equilateral triangle. This result makes use of sub-problem 1 and simple symmetry arguments. 
The third sub-problem is to determine the probability that lines defined by $\Re[\Delta] = 0$ and $\Im[\Delta] = 0$ additionally intersect within the triangle,.
The results of sub-problems 2 and 3 are then combined to give the final result for $\rho_\text{VV}$.

\subsection{Sub-problem 1: Probability of observing $\Re[\Delta] = 0$ or $\Im[\Delta] = 0$ on a 1-dimensional line segment}

\begin{figure}[t]
\begin{center}
\includegraphics[width=0.3\textwidth]{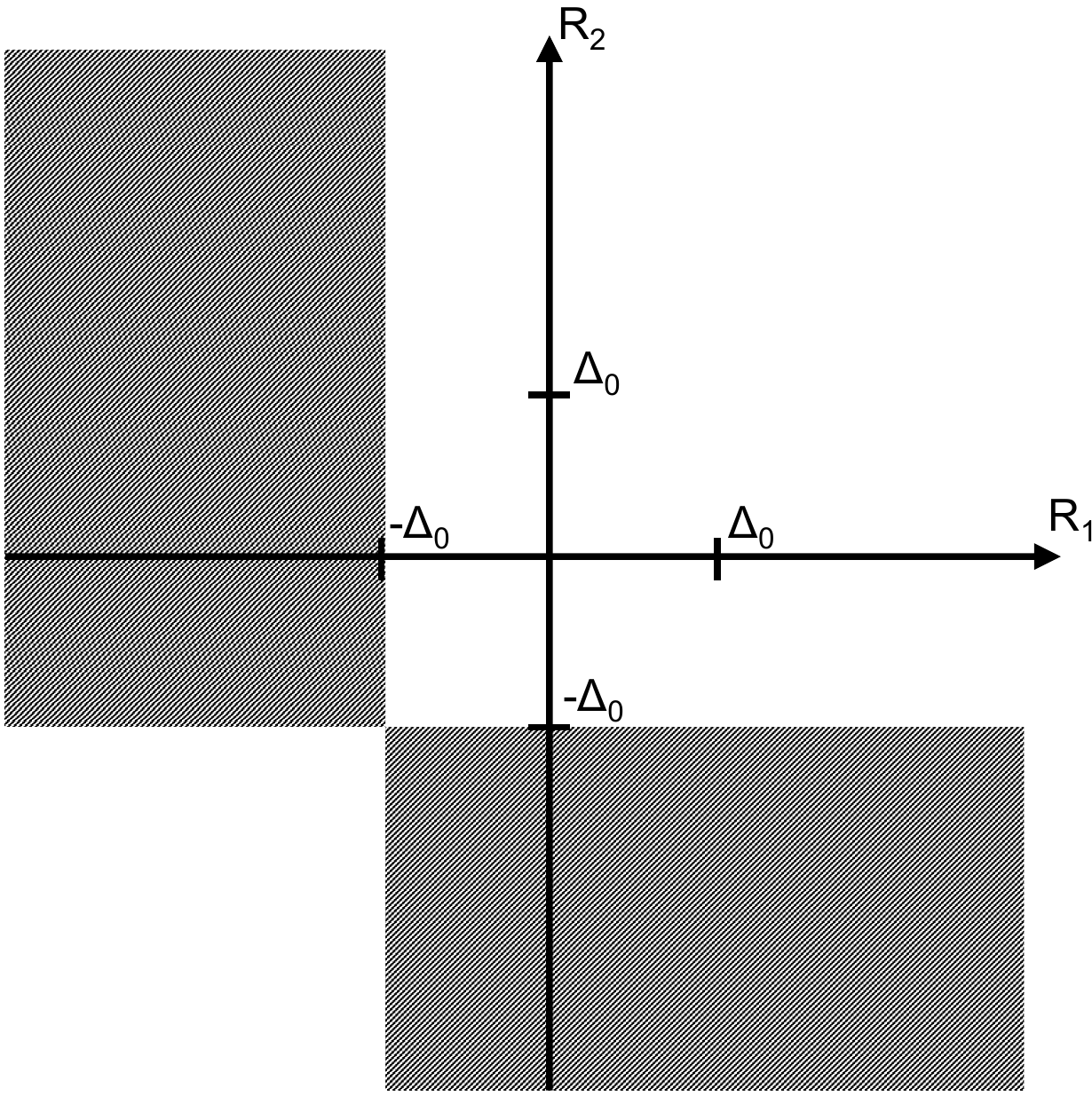}
\end{center}
\vspace{-0.5cm}
\caption{The parameter space of the real component $R(x,y)$ of the fluctuating part of the inter-valley coupling $\Delta(x,y)$,
for the two points $(x,y)= {\mathbf r}_1$ and ${\mathbf r}_2$. Here, $R_1 = R({\mathbf r}_1)$ and $R_2 = R({\mathbf r}_2)$. The hatched regions denote the parameter space regions where $\Re[\Delta({\mathbf r}_1)]$ and $\Re[\Delta({\mathbf r}_2)]$ are of opposite sign.}
\label{FIGS2}
\vspace{-1mm}
\end{figure}

We define $P_{\bullet\!-\!\bullet}^{R}(s)$ as the probability that $\Re[\Delta(x,y)] = \Delta_0 + R(x,y) = 0$ somewhere along a line segment of length $s$ as $s \rightarrow 0$.
In this limit, $P_{\bullet\!-\!\bullet}^{R}(s)$ is equal to the probability that $\Re[\Delta({\mathbf r}_1)]$ and $\Re[\Delta({\mathbf r}_2)]$ are of opposite sign, where $\BS{r}_1$ and $\BS{r}_2$ are the endpoints of the line segment.
Within the parameter space of $R_1 = R({\mathbf r}_1)$ and $R_2 = R({\mathbf r}_2)$, this condition is satisfied in the hatched regions of Fig.~\ref{FIGS2}. 

The covariance properties of $\Delta(x,y)$ are known~\cite{Losert2023}, with the covariance matrix of $R_1$ and $R_2$ given by 
\begin{equation}
    \Sigma = \frac{\sigma_{\Delta}^2}{2}
    \begin{pmatrix}
    1 & e^{-s^2/(2 \ell_t^2)} \\
    e^{-s^2/(2 \ell_t^2)} & 1
    \end{pmatrix}.
\end{equation}
The corresponding joint probability density function is then given by the multi-variable normal distribution
\begin{equation}
    f_{R_{1}, R_{2}}(R_1, R_2) = 
    \frac{1}{2 \pi\sqrt{|\Sigma|}}
    e^{-\frac{1}{2}{\mathbf R}^T\Sigma^{-1}{\mathbf R}},
\end{equation}
where ${\mathbf R} = (R_1,R_2)^T$. 
$P_{\bullet\!-\!\bullet}^{R}(s)$ is then given by the integral
\begin{equation}
    P_{\bullet\!-\!\bullet}^{R}(s) = 2 
    \int_{- \infty}^{-\Delta_0} \int_{-\Delta_0}^{\infty} 
    f_{R_{1}, R_{2}}(R_1, R_2) \, dR_1 dR_2 + \mathcal{O}(s^2), \label{Int1}
\end{equation}
which simplifies via a change of variables:
\begin{align}
    R_{1}^\prime &= \frac{R_1 + R_2}{\sqrt{2}}, \\
    R_{2}^\prime &= \frac{-R_1 + R_2}{\sqrt{2}}.
\end{align}
This diagonalizes the covariance matrix, giving 
\begin{equation}
    \Sigma^\prime = \frac{\sigma_{\Delta}^2}{2}
    \begin{pmatrix}
    1 + e^{-s^2/(2 \ell_t^2)} & 0  \\
    0 & 1 - e^{-s^2/(2 \ell_t^2)}
    \end{pmatrix} 
    \approx 
    \frac{\sigma_{\Delta}^2}{2}
    \begin{pmatrix}
    2& 0  \\
    0 & \frac{s^2}{2\ell_t^2}
    \end{pmatrix},
\end{equation}
and the factorizable joint probability distribution function
\begin{equation}
    f_{R_{1}^\prime, R_{2}^\prime}(R_1^\prime, R_2^\prime) = f_{R_{1}^\prime}(R_1^\prime) f_{R_{2}^\prime}(R_2^\prime)
    \approx 
    \frac{1}{2 \pi\sigma_{\Delta}^2} \left(\frac{2\ell_t}{s}\right)
    \exp[-\frac{(R_1^\prime)^2}{2\sigma_{\Delta}^2}]
    \exp[-\frac{2\ell_t^2}{s^2}\frac{(R_2^\prime)^2}{2\sigma_{\Delta}^2}]
    .
\end{equation}
The integral in Eq.~(\ref{Int1}) then reduces to
\begin{equation}
    P_{\bullet\!-\!\bullet}^{R}(s) = \frac{1}{\pi \sigma_{\Delta}^2} \left(\frac{2 \ell_t}{s}\right) 
    \int_{0}^{\infty}
    \exp[-\frac{2\ell_t^2}{s^2}\frac{(R_2^\prime)^2}{2\sigma_{\Delta}^2}]
    \left(
    \int_{-\sqrt{2}\Delta_0 - R_2^\prime}^{-\sqrt{2}\Delta_0 + R_2^\prime} 
    \exp[-\frac{(R_1^\prime)^2}{2\sigma_{\Delta}^2}]    
    \, dR_1^\prime
    \right)  
    \, dR_2^\prime
    + \mathcal{O}(s^2). \label{Int2}
\end{equation}
Here, the inner $R_1^\prime$ integral can be written in terms of an error function involving $R_2^\prime$. 
Only small values of $R_2^\prime$ are relevant in the limit $s \rightarrow 0$, however, because the first exponential factor in Eq.~(\ref{Int2}) is strongly suppressed.
The inner $R_1^\prime$ integral can therefore be expanded to first order in $R_2^\prime$, giving
\begin{equation}
    \int_{-\sqrt{2}\Delta_0 - R_2^\prime}^{-\sqrt{2}\Delta_0 + R_2^\prime} 
    \exp[-\frac{(R_1^\prime)^2}{2\sigma_{\Delta}^2}]    
    \, dR_1^\prime = 
    2 R_2^\prime e^{-\Delta_0^2 / \sigma_{\Delta}^2} +
    \mathcal{O}((R_2^\prime)^3). \label{Int3}
\end{equation}
Thus, Eq.~(\ref{Int2}) is given by
\begin{equation}
    \begin{split}
    P_{\bullet\!-\!\bullet}^{R}(s) =& \frac{4}{\pi \sigma_{\Delta}^2} \left(\frac{\ell_t}{s}\right) 
    e^{-\Delta_0^2 / \sigma_{\Delta}^2}
    \int_{0}^{\infty}
    R_2^\prime
    \exp[-\frac{2\ell_t^2}{s^2}\frac{(R_2^\prime)^2}{2\sigma_{\Delta}^2}] 
    \, dR_2^\prime
    + \mathcal{O}(s^2)  \\
    =& \frac{s}{\pi \ell_t} e^{-\Delta_0^2 / \sigma_{\Delta}^2} + \mathcal{O}(s^2),
    \end{split}
\end{equation}
giving the final solution to sub-problem 1.

Following similar methods, we obtain the probability $P_{\bullet\!-\!\bullet}^{I}(s)$ of finding $\Im[\Delta(x,y)] = 0$ somewhere along a line segment of length $s$, yielding 
\begin{equation}
    P_{\bullet\!-\!\bullet}^{I}(s) = \frac{s}{\pi \ell_t}  + \mathcal{O}(s^2)
\end{equation}

\subsection{Sub-problem 2: Probability that the lines $\Re[\Delta] = 0$ or $\Im[\Delta] = 0$ cut through an equilateral triangle}

We define $P_{\triangle}^{R}(s)$ as the probability that a line defined by $\Re[\Delta] = 0$ passes through an equilateral triangle of side-length $s$, as $s \rightarrow 0$.
In this limit, any smooth line would have to pass through two of the three edges of the triangle.
Due to rotational symmetry, the probability of a $\Re[\Delta] = 0$ line passing through any of the three triangle edges is the same.
We therefore have
\begin{equation}
\begin{split}
    \lim_{s \rightarrow 0} P_{\triangle}^{R}(s) =&  \frac{3}{2}\lim_{s \rightarrow 0} P_{\bullet\!-\!\bullet}(s) \\
    =&
    \frac{3s}{2\pi \ell_t} e^{-\Delta_0^2 / \sigma_{\Delta}^2}.
    \end{split} \label{PtriR}
\end{equation}

Similarly, the probability $P_{\triangle}^{I}(s)$ of a line defined by $\Im[\Delta] = 0$ passing through an equilateral triangle of side-length $s$, as $s \rightarrow 0$, is given by
\begin{equation}
    \lim_{s \rightarrow 0} P_{\triangle}^{I}(s)
    =
    \frac{3s}{2\pi \ell_t} . \label{PtriI}
\end{equation}
Finally, since $R(\BS{r})$ and $I(\BS{r})$ are statistically independent, the probability $P_{\triangle}(s)$ of lines defined by $\Re[\Delta] = 0$ and $\Im[\Delta] = 0$ both passing through an equilateral triangle of side-length $s$ is given by the simple product
\begin{equation}
\begin{split}
    P_{\triangle}(s) =& P_{\triangle}^{R}(s) P_{\triangle}^{I}(s) \\
    =& 
    \frac{9s^2}{4\pi^2 \ell_t^2} e^{-\Delta_0^2 / \sigma_{\Delta}^2} + \mathcal{O}(s^3).
    \end{split} \label{Ptri}
\end{equation}

\subsection{Sub-problem 3: Probability that lines $\Re[\Delta] = 0$ and $\Im[\Delta] = 0$ cross inside an equilateral triangle}

We define $P_\text{int}$ as the probability that lines defined by $\Re[\Delta] = 0$ and $\Im[\Delta] = 0$ cross inside an equilateral triangle, given that both lines pass through the triangle.
To compute $P_\text{int}$, we must specify the distributions from which the lines are sampled. 
In the limit $s\rightarrow 0$, smooth lines passing through an equilateral triangle are effectively straight line segments.
The two lines pass through at least one common edge.
Without loss of generality, we choose this edge to be the bottom edge of the triangle, along the $x$-axis, between the points $x = \pm s/2$. 
Figure~\ref{FIGS1} shows an example configuration, with $\Re[\Delta] = 0$ and $\Im[\Delta] = 0$ illustrated as red and blue line segments, respectively.
Since the dots used to compute $\Delta$ are isotropic, the orientations of the two lines ($\theta_R$ and $\theta_I$) should have no preferred direction and should be drawn from a uniform distribution over $[0,\pi)$.
Moreover, due to translational invariance, the points of intersection between the two lines and the bottom edge of the triangle, defined as $x_R$ and $x_I$, are drawn from a uniform distribution between $x=\pm s/2$. 
The $x$ locations where the lines exit the triangle are denoted as $x_{R,f}$ and $x_{I,f}$ in the figure.
We see that the two lines intersect if and only if $x_{R,f} - x_{I,f}$ and $x_{R} - x_{I}$ are of opposite sign.
The probability of intersection is therefore given by 
\begin{equation}
\begin{split}
    P_\text{int} =& ~
    \frac{1}{4\pi^2 s^2}
    \int_{-\pi}^{\pi}
    \int_{-\pi}^{\pi}
    \int_{-s/2}^{x_I}
    \int_{-s/2}^{s/2}
    \Theta\left[
    x_{R,f}(x_R,\theta_R) - x_{I,f}(x_I,\theta_I)
    \right] \, dx_I dx_R d\theta_I d\theta_R\\
    &+ 
    \frac{1}{4\pi^2 s^2}
    \int_{-\pi}^{\pi}
    \int_{-\pi}^{\pi}
    \int_{x_I}^{s/2}
    \int_{-s/2}^{s/2}
    \Theta\left[
    -x_{R,f}(x_R,\theta_R) + x_{I,f}(x_I,\theta_I)
    \right] \, dx_I dx_R d\theta_I d\theta_R .
    \end{split}
\end{equation}
Numerical evaluation of the integral yields
\begin{equation}
    P_\text{int} \approx 0.349. \label{Pint}
\end{equation}

\subsection{Final result}
Putting everything together, we see that the probability $P_1^\triangle(s)$ that an equilateral triangle of side-length $s$ encloses a single VV, as $s \rightarrow 0$, is given by the probability of having the lines $\Re[\Delta] = 0$ and $\Im[\Delta] = 0$ both traverse the triangle, multiplied by the probability $P_\text{int} $ that they also cross within the triangle:
\begin{equation}
    \lim_{s \rightarrow 0} P_1^\triangle(s) = P_\text{int}  \lim_{s \rightarrow 0} P_\triangle(s). \label{P_1Tri}
\end{equation}
Combining Eqs.~(\ref{rhoVV1}), (\ref{Ptri}), (\ref{Pint}), and (\ref{P_1Tri}), the density of VVs is finally given by
\begin{equation}
    \begin{split}
        \rho_\text{VV} =& P_\text{int} \frac{9}{\sqrt{3}\pi^2 \ell_t^2} e^{-\Delta_0^2 / \sigma_{\Delta}^2} \\
        &\approx \frac{0.184}{\ell_t^2}e^{-\Delta_0^2 / \sigma_{\Delta}^2},
    \end{split}
\end{equation}
which is the result given in Eq.~(5) of the main text.

\section{Details of the derivation of the model Hamiltonian}

In this Supplementary Materials section, we provide details regarding the calculation the effective Hamiltonian Eq.~(3) in the main text. 
We use the same effective mass model as \cite{Woods2026a}. 
For clarity, we summarize some of those calculations here. 
We emphasize that this effective model yields spin-orbit coefficients, $g$-factors, and valley splittings that are good agreement with quantitatively accurate $sp^3d^5s^*$ tight-binding models \cite{Woods2023a,Losert2023}.

We consider a quantum dot in a Si/SiGe quantum well with a sigmoidal Ge concentration profile
\begin{equation}
    n_{\T{Ge}}(z) = 
    \frac{n_{\T{bar}}}
    {1 + e^{4z/W}}, \label{GeProfile}
\end{equation}
where $n_{\T{bar}} = 0.3$ is the Ge concentration in the SiGe barrier region and $W = 0.8~\T{nm}$ is the interface width measured in state-of-the-art devices~\cite{Wuetz2021}.
We assume a vertical electric field of $F_z = 5~\T{mV/nm}$ that pulls the electronic wavefunction up against the quantum well interface. 
[Note that a second SiGe barrier region is excluded from Eq.~(\ref{GeProfile}), as the vertical electric field is large enough that a second barrier would insignificantly affect the wavefunction, except for cases of very thin quantum wells.]
In addition, we assume an isotropic in-plane harmonic, quantum-dot confinement potential with an orbital spacing of $\hbar \omega_t = 2~\T{meV}$.

The system is modeled by an effective-mass Hamiltonian with both spin and valley degrees of freedom, given by
\begin{equation}
    \begin{split}
        H =& 
        \frac{\hbar^2}{2} 
    \left(
    \frac{\hat{\pi}_z^2}
    {m_l} + \frac{\hat{k}_x^2 + \hat{k}_y^2}{m_t}
    \right)
    + V({\mathbf r}) 
    + \frac{\mu_B g_0}{2}  {\mathbf B} \cdot \BS{\sigma}
    + \left(
    \left[
    V({\mathbf r}) 
    e^{-i 2k_0 z} + \beta_0 e^{i 2k_1 z}
    \left(\hat{k}_x \sigma_x - \hat{k}_y \sigma_y\right)
    \right] \tau_- + \text{h.c.}
    \right),
    \end{split} \label{Ham}
\end{equation}
where $\hat{\pi}_z \! = \! \hat{k}_z + (e/\hbar)A_z(x,y)$,
$\hat{k}_j \! = \! -i\partial_j$, $m_l\! =\! 0.91m_e$ and $m_{t}\! =\!0.19m_e$~\cite{Zwanenburg2013} are the longitudinal and transverse effective masses, respectively. 
Note that $\BS{A} = A_z(x,y) \hat{z}$ with $A_z(x,y) = B_x y - B_y x$ is the vector potential satisfying $\BS{B} = \nabla \times \BS{A}$.
Here, $\sigma_j$ are Pauli spin matrices, with $j \in \{x,y,z\}$, and $\tau_{\pm} = \ket{\pm z}\bra{\mp z}$ is a valley raising/lowering operator with $\pm z$ denoting the two valleys near the $Z$ point in the conduction band structure of Si. 
The confinement potential is given by 
\begin{equation}
    V({\mathbf r})  = 
    V_l(z) 
    + \frac{m_t \omega_t^2}{2} \left(x^2 + y^2\right)
    + V_{\T{dis}}({\mathbf r}) , \label{POT}
\end{equation}
where $V_l$ is the quantum well confinement potential arising from the Ge concentration profile $n_{\T{Ge}}(z)$ and the external electric field, but does not include random-alloy fluctuations,
$\hbar \omega_t$ is the orbital splitting of an in-plane harmonic confinement potential representing a quantum dot, 
and $V_{\T{dis}}$ describes the potential fluctuations caused by random-alloy disorder, which are discussed in more detail below.
The longitudinal potential is explicitly given by
\begin{equation}
    V_l(z) = eF_z z + E_{\T{Ge}}n_{\T{Ge}}(z),
\end{equation}
where $E_{\T{Ge}} = 0.6~\T{eV}$ gives the desired band offset between atomic layers with different Ge concentrations.
The $\tau_-$ term in Eq.~(\ref{Ham}) describes inter-valley coupling, where the $2k_0$ and $2k_1$ terms describe normal (i.e. spin-conserving) valley coupling and Dresselhaus spin-orbit coupling, respectively. Here, $2k_0 \approx 0.83 (4\pi/a_{0})$ and $2k_1 = 4\pi/a_{0} - 2k_0$ correspond to wave vectors connecting valleys in the same or neighboring Brillouin zones, respectively, in the Si band structure.
We emphasize that these inter-valley coupling terms yield precisely the same momentum selection rules as $sp^3d^5s^*$ tight-binding models~\cite{Woods2023a}. 
In other words, the spin-orbit coupling terms in the effective-mass Hamiltonian are not ad-hoc, but rather, they can be related to the physics of more-detailed tight-binding models.

The Zeeman coupling, vector-potential coupling, inter-valley coupling, and alloy-disorder terms are all small compared to the kinetic energy and confinement terms in Eq.~(\ref{Ham}).
We therefore work in the basis that diagonalizes the Hamiltonian in the absence of such perturbations.
These basis states are denoted by $\ket{\nu,n_x,n_y,\tau,\sigma}$
where $\nu$, $n_x$, and $n_y$ are spatial orbital indices, and $\tau \in \{+z,-z\}$ and $ \sigma \in \{\uparrow,\downarrow\}$ denote the valley and spin, respectively. 
The spatial wavefunctions are given by $\bra{\BS{r}}\ket{\nu,n_x,n_y} = \varphi_{\nu}(z) \chi_{n_x,n_y}(x,y)$,
where $\chi_{n_x,n_y}$ is an in-plane harmonic-oscillator orbital with energy $\hbar\omega_t(n_x\! +\! n_y\!+\!1)$ arising from the quantum-dot harmonic confinement potential, and
$\varphi_{\nu}$ is a subband wavefunction satisfying 
\begin{equation}
    \left[\frac{\hbar^2 \hat{k}_z^2}{2m_{t}} + V_l(z)\right]\varphi_{\nu}(z) = \varepsilon_\nu \varphi_{\nu}(z), \label{SubbandEq}
\end{equation}
where $\varepsilon_{\nu}$ is the subband energy. 
Here, we solve for the subband wavefunctions $\varphi_{\nu}$ and energies $\varepsilon_{\nu}$ numerically using conventional finite-difference methods for one-dimensional differential equations.

In the new basis, the Hamiltonian is given by
\begin{equation}
    \begin{split}
    &H^{\mu,\nu}_{{\mathbf m},{\mathbf n}}
    =
    ~\delta_{\mu,\nu} \delta_{{\mathbf m},{\mathbf n}}
    \left[
    \varepsilon_{\nu,\BS{n}}
    + \frac{\mu_B g_0}{2} {\mathbf B} \cdot \BS{\sigma}
    \right] 
    + k_{z}^{\mu, \nu} \mathcal{A}_{{\mathbf m},{\mathbf n}}
    + \left( 
    \left[
    \tilde{\Delta}_{{\mathbf m},{\mathbf n}}^{\mu,\nu} + \beta_{\mu, \nu}
    D_{{\mathbf m},{\mathbf n}}
    \right] \tau_-
    + \text{h.c.} \right)
    + \mathcal{V}_{{\mathbf m},{\mathbf n}}
    ,
    \end{split} \label{HSubband}
\end{equation}
where $H_{{\mathbf m},{\mathbf n}}^{\mu,\nu} = \mel{\mu,{\mathbf m}}{H}{\nu,{\mathbf n}}$ and ${\mathbf n} = (n_x,n_y)$. 
Here, the various matrix elements are given by 
\begin{align}
    & \varepsilon_{\nu,{\mathbf n}} = \varepsilon_{\nu} + \hbar \omega_t\left(n_x + n_y\right), \\
    &k_z^{\mu, \nu} = \mel{\varphi_{\mu}}{\hat{k}_z}{\varphi_{\nu}} \in i\mathbb{R}, \label{kz} \\
    &\mathcal{A}_{{\mathbf m},{\mathbf n}} = 
    \frac{\hbar^2 \pi}{m_l \Phi_0} \mel{\chi_{\mathbf m}}{A_z}{\chi_{\mathbf n}} =
    \frac{\hbar^2 \pi}{m_l \Phi_0}\left(B_x y^{{\mathbf m},{\mathbf n}}- B_y x^{{\mathbf m},{\mathbf n}}\right) \in \mathbb{R}
    ,
    \\
    &\tilde{\Delta}_{{\mathbf m},{\mathbf n}}^{\mu,\nu} = 
    \mel{\varphi_{\mu} \chi_{\mathbf m}}{V e^{-i 2k_0 z}}{\varphi_{\nu}\chi_{\mathbf n}} \in \mathbb{C}, \label{DeltaMtxElem}\\
    &\beta_{\mu, \nu} = \beta_0 \mel{\varphi_{\mu}}{e^{i 2 k_1 z}}{\varphi_{\nu}} \in \mathbb{C}, \label{betaMuNu}\\
    &D_{{\mathbf m},{\mathbf n}} = k_x^{{\mathbf m},{\mathbf n}}\sigma_x - k_y^{{\mathbf m},{\mathbf n}} \sigma_y, \\
    &\mathcal{V}_{{\mathbf m},{\mathbf n}}^{\mu,\nu} = 
    \mel{\varphi_{\mu} \chi_{\mathbf m}}{V_{\T{dis}}}{\varphi_{\nu}\chi_{\mathbf n}} \in \mathbb{R}, \label{VmtxElem}
\end{align}
where
\begin{align}
    x^{\mathbf{m},\mathbf{n}} &= \delta_{m_y,n_y}\frac{\ell_t}{\sqrt{2}}\left( \sqrt{m_x}\delta_{m_x,n_x+1} + \sqrt{n_x}\delta_{m_x,n_x-1}\right) \in \mathbb{R},\\
    y^{\mathbf{m},\mathbf{n}} &= \delta_{m_x,n_x}\frac{\ell_t}{\sqrt{2}}\left( \sqrt{m_y}\delta_{m_y,n_y+1} + \sqrt{n_y}\delta_{m_y,n_y-1}\right) \in \mathbb{R},\\
    k_x^{\mathbf{m},\mathbf{n}} &= \delta_{m_y,n_y}\frac{i}{\sqrt{2} \ell_t}\left( \sqrt{m_x}\delta_{m_x,n_x+1} - \sqrt{n_x}\delta_{m_x,n_x-1}\right) \in i\mathbb{R},\\ 
    k_y^{\mathbf{m},\mathbf{n}} &= \delta_{m_x,n_x}\frac{i}{\sqrt{2} \ell_t}\left( \sqrt{m_y}\delta_{m_y,n_y+1} - \sqrt{n_y}\delta_{m_y,n_y-1}\right) \in i\mathbb{R},
\end{align}
and $\ell_t = \sqrt{\hbar/( m_t \omega_t)}$ is the characteristic dot size in the transverse direction. 

To obtain the effective Hamiltonian Eq.~(3) in the main text, we perform a second-order Schrieffer-Wolff transformation of Eq.~(\ref{HSubband}), treating all the excited orbital states $\nu,{\mathbf n}>0$ as perturbations.
In this way, the inter-valley coupling is given by~\cite{Woods2026a}
\begin{equation}
    \Delta = \tilde{\Delta}_{\BS{0},\BS{0}}^{0,0} -2
    {\sum_{\nu,{\mathbf n}}}' \frac{
    \mathcal{V}^{0,\nu}_{\BS{0},{\mathbf n}} \tilde{\Delta}^{\nu,0}_{{\mathbf n},\BS{0}}
    }{\varepsilon_{\nu, {\mathbf n}} - \varepsilon_{0}} , \label{DeltaSup}
\end{equation}
where the prime notation indicates that the ground state is not included in the sum.
The spin-orbit coupling term is given by
\begin{equation}
    g_{\tau} =  
    \frac{2e \hbar }{\mu_B m_l}
    \sum_{\nu >0}\frac{(ik_z^{0,\nu}) \beta_{\nu,0}}{\varepsilon_{\nu} - \varepsilon_{0} + \hbar \omega_t}  , \label{gtau0}
\end{equation}
which involves momentum and spin-orbit matrix elements defined in Eqs.~(\ref{kz}) and (\ref{betaMuNu}).
We note that, at this level of approximation, $\Delta$ includes correction terms arising from the alloy-disorder matrix elements $\mathcal{V}$ and $\tilde{\Delta}$, which couple the lowest-energy orbital to higher-energy orbitals, while $g_\tau$ does not.
To compute these disorder terms, we make use of their statistical description, given by a multivariate normal distribution whose covariance elements are derived in the Supplementary Materials of Ref.~\cite{Woods2026a}.
Alloy disorder is therefore included by calculating the relevant covariance elements and sampling the resulting multivariate normal distributions.

To calculate the $g$-factor numerically, as in necessary for generating the data in Fig. 2(g) of the main text, we project Eq. (\ref{HSubband}) onto a finite basis set and solve for the two lowest-energy eigenstates in the limit of $B \rightarrow 0$. We check for convergence by increasing the number of subbands and harmonic oscillator orbitals included in the basis set.

\section{Explanation for why higher-order alloy-disorder induced fluctuations in $g$ are small}

In the main text, we noted that some higher-order alloy-disorder terms were not included in the effective Hamiltonian of Eq.~(3),
although in principle, they could interfere with our ability to invert $g$ to obtain the valley phase $\phi_v$.
In Fig.~2(g), we showed that such corrections to $g$ are very small by collecting statistics on the valley-phase error $|{\phi}_{v}^{(g)}| - |\phi_{v}|$. 
Here, we provide two arguments for why the disorder-induced corrections are small.

First, the corrections to $g_\tau$ involving disorder enter only at third-order in the perturbation theory, while second-order corrections are all deterministic. 
Compared to second-order terms, the disorder-induced terms are suppressed by the ratio of the disorder energy scale (tens of $\mu$eV) to the subband energy splitting (tens of meV), which is very small.
Second, the deterministic contributions to $g_\tau$ are seen to be relatively large compared to the deterministic contributions to $\Delta_0$. 
This is because the spin-orbit matrix element $\beta_{\mu,\nu}$ in Eq.~(\ref{betaMuNu}) requires ``sharp'' features in the quantum well, of length scale $\pi/k_1 \approx 1.6~\T{nm}$, which are fairly easy to engineer in modern devices.
This should be compared to the inter-valley coupling term $\tilde{\Delta}_{{\mathbf m},{\mathbf n}}^{\mu,\nu}$ in Eq.~(\ref{DeltaMtxElem}) that requires sharp features of length scale $\pi/k_0 \approx 0.32~\T{nm}$, which are much more difficult to achieve in experiments, as noted in several previous works~\cite{Losert2023,Lima2023a,Lima2023b,Pena2024}.
It is therefore reasonable that $\Delta$ is typically dominated by disorder effects, while $g_{\tau}$ is most strongly affected by deterministic effects.

\end{document}